\def\ra{\rangle}
\def\la{\langle}

\def\Cb{\mathbb{C}}

\def\be{\begin{equation}}
\def\ee{\end{equation}}
\def\ba{\begin{array}}
\def\ea{\end{array}}

\def\Nb{{I\!\! N}}

\def\Cb{{\Bbb C}}

\def\beq{\begin{equation}}
\def\eeq{\end{equation}}
\def\bea{\begin{eqnarray}}
\def\eea{\end{eqnarray}}
\def\ba{\begin{array}}
\def\ea{\end{array}}

\documentstyle[12pt]{article}
\input amssym.def
\def\ra{\rangle}
\def\la{\langle}

\begin{document}
\baselineskip18pt
\begin{center}
{\LARGE \bf  Canonical Form and Separability of PPT}\\
\medskip
{\LARGE \bf States on Multiple Quantum Spaces}
\end{center}
\vskip 2mm

\begin{center}
{\normalsize Xiao-Hong Wang$^1$ and Shao-Ming Fei$^{1,\ 2}$}
\end{center}

\begin{center}

\begin{minipage}{4.8in}
{\small \sl $^1$
Department of Mathematics, Capital  Normal University, Beijing,
China}

{\small \sl $^2$ Institute of Applied Mathematics,
University of Bonn,  53115 Bonn, Germany}
\end{minipage}
\end{center}
\vskip 3mm

\begin{center}
\begin{minipage}{5in}
{\bf Abstract}
 By using the ``subtracting projectors" method in proving the
separability of PPT states on multiple quantum spaces, we derive a
canonical form of PPT states in ${\Cb}^{K_1} \otimes {\Cb}^{K_2}
\otimes \cdots \otimes {\Cb}^{K_m} \otimes {\Cb}^N$ composite
quantum systems with rank $N$, from which a sufficient
separability condition for these states is presented.
\end{minipage}
\end{center}
\vskip 4mm

As the key resource in quantum information processing
\cite{bennet98}, quantum entanglement has resulted in the
explosion of interest in quantum computing and communication in
recent years \cite{book}. The separability of quantum pure states
is well understood \cite{peresbook}. However for mixed states,
although there are already many results related to the
separability criterion, the physical character and mathematical
structure of the quantum entanglement are still far from being
satisfied. The PPT (positive partial transpose) criteria play very
important roles in the investigation of separability. It is of
significance to investigate the general form of PPT states to
study the separability and bound entangled states. In \cite{22n}
the PPT (mixed) states on $\Cb^2\times \Cb^2 \times \Cb^N$ with
rank $N$ has been studied and a separability criterion has been
obtained. These results have been generalized to the case of PPT
states on $\Cb^2\times \Cb^3 \times \Cb^N$ with rank $N$
\cite{23n}, and the case of high dimensional tripartite systems
\cite{KMN} and $\Cb^2\times \Cb^2\times \Cb^2 \times \Cb^N$
\cite{222n}. In this note we summarize our efforts in striving at
some understanding of the properties of quantum entanglement for
composite systems. We generalize the results to the most general
case of the canonical form of PPT states in ${\Cb}^{K_1} \otimes
{\Cb}^{K_2} \otimes \cdots \otimes {\Cb}^{K_m} \otimes {\Cb}^N$
composite quantum systems with rank $N$, and study the
separability condition for these states in terms of the canonical
form.

Let ${\Cb}^K$ be $K$-dimensional complex vector space with orthonormal
basis $\{\bf \vert i\rangle\}$, $i=0,...,K-1$. A general pure
state on ${\Cb}^{K_1}\otimes {\Cb}^{K_2}\otimes\dots\otimes
{\Cb}^{K_m}$ is of the form
\begin{equation}
\label{16} |\Psi\ra=\sum_{i=0}^{K_1-1}\sum_{j=0}^{K_2-1}\dots
\sum_{k=0}^{K_m-1} a_{ij\dots k}|i,j,\dots, k\ra,
\end{equation}
where $|i,j,\dots, k\ra=|i\ra\otimes |j\ra\otimes\dots\otimes
|k\ra$, $a_{ij\dots k}\in {\Bbb C}$, $\sum a_{ij\dots k}a_{ij\dots
k}^\ast=1$ ($\ast$ denoting complex conjugation). $|\Psi\ra$ is
said to be (fully) separable if $a_{ij\dots k}=a_{i}a_{j}...a_{k}$
for some $a_{i},~a_{j},~...,~a_{k}\in\Cb$. A mixed state on
${\Cb}^{K_1}\otimes {\Cb}^{K_2}\otimes\dots\otimes {\Cb}^{K_m}$ is
described by a density matrix $\rho$,
\begin{equation}
\label{rho1} \rho=\sum_{i=1}^M p_i |\Psi_i\ra\langle\Psi_i\vert,
\end{equation}
for some $M\in\Nb$, $0<p_i\leq 1$, $\sum_{i=1}^M p_i=1$,
$|\Psi_i\ra$s are pure states of the form (\ref{16}) and
$\langle\Psi_i\vert$ is the transpose and conjugation of
$|\Psi_i\ra$.
We call state $\rho$ PPT if $\rho^{T_l}\geq 0$, $\forall l$,
where $\rho^{T_l}$ is the transpose of $\rho$ with respect to
the $l$-th subspace.

In the following we denote by $R(\rho)$, $K(\rho)$, $r(\rho)$ and
$k(\rho)$ the range, kernel, rank, dimension of the kernel of
$\rho$ respectively, where, by definition $K(\rho)=
\{|\phi\ra:\rho|\phi\ra=0\}$, $R(\rho)= \{|\phi\ra:\exists
|\psi\ra,$ such that $ |\phi\ra=\rho|\psi\ra\}.$

We consider now composite quantum systems in ${\Cb}^{K_1}_{\sc
A_1} \otimes {\Cb}^{K_2}_{\sc A_2} \otimes \cdots \otimes
{\Cb}^{K_m}_{{\sc A}_m} \otimes {\Cb}^N_{{\sc A}_{m+1}}$  with
$r(\rho)=N$, where ${\sc A}_i$ denotes the $i$-th subsystem, $K_i$
stands for the dimension of the $i$-th complex vector space,
$m,\,N\in\Nb$.

We first derive a canonical form of PPT states in
${\Cb}^2_{\sc A_1} \otimes {\Cb}^2_{\sc A_2} \otimes \cdots \otimes
{\Cb}^2_{{\sc A}_m} \otimes {\Cb}^N_{{\sc A}_{m+1}}$ with rank
$N$, which allows for an explicit decomposition of a given state
in terms of convex sum of projectors on product vectors. Let
$|0_{{\sc A}_1}\ra$, $|1_{{\sc A}_1}\ra$; $|0_{{\sc A}_2}\ra$,
 $|1_{{\sc A}_2}\ra$; $\cdots $; $|0_{{\sc A}_m}\ra$, $|1_{{\sc
A}_m}\ra$ and $|0_{{\sc A}_{m+1}}\ra\ , \cdots \ ,|(N-1)_{{\sc
A}_{m+1}}\ra$ be some local bases of the sub-systems ${\sc
A}_1,~{\sc A}_2,\cdots, ~{\sc A}_m$, and ${\sc A}_{m+1}$
respectively. In terms of the method used in \cite{22n,23n,KMN,222n},
we have

{\bf Lemma 1. }\ \  Every PPT state $\rho$ in ${\Cb}^2_{\sc A_1}
\otimes {\Cb}^2_{\sc A_2} \otimes \cdots \otimes {\Cb}^2_{{\sc
A}_m} \otimes {\Cb}^N_{{\sc A}_{m+1}}$ such that $$r(\la 1_{A_1},
1_{A_2},\cdots, 1_{A_m}|\rho |1_{A_1}, 1_{A_2},\cdots,
1_{A_m}\ra)=r(\rho)=N,$$ can be transformed into the following
canonical form by using a reversible local operation:
\be\label{lemma1} \rho=\sqrt{F}T^{\dag}T \sqrt{F}, \ee where
$T=(D_m\ \ I) \otimes (D_{m-1}\ \ I) \otimes \cdots \otimes (D_1\
\ I)$, $D_i$, $F$ and the identity $I$ are $N\times N$ matrices
acting on ${\Cb}_{{\sc A}_{m+1}}^N$ and satisfy the following
relations: $[D_i,\ D_j]=[D_i,\ {D_j}^{\dag}]=0$, and $F=F^{\dag}$
($\dag$ stands for the transpose and conjugate), $i,j=1,2,\cdots \
m$.

Using Lemma 1 we can prove the the following Theorem:

{\bf Theorem 1.}\ \ A PPT-state $\rho$ in ${\Cb}^2 \otimes {\Cb}^2
\otimes \cdots \otimes {\Cb}^2 \otimes {\Cb}^N$ with $r(\rho)=N$
is separable if there exists a product basis $|e_{A_1},\
e_{A_2},\cdots , \ e_{A_m}\ra$ such that $$r(\la e_{A_1},\
e_{A_2},\cdots, \ e_{A_m}|\rho|e_{A_1},\ e_{A_2}, \cdots, \
e_{A_m} \ra)=N$$.

{\bf Proof. }\ \ According to Lemma 1 the PPT state $\rho$ can be
written as the form of (\ref{lemma1}). Since all the $D_i$ and
$D_j^\dag$ commute, they have common eigenvectors $|f_n\ra$. Let
$a^n_1$, $a^n_2$, $\cdots$, $a^n_m$  be the corresponding
eigenvalues of $D_1 ,~ D_2,~ \cdots,~ D_m $ respectively. We have
$$
\begin{array}{rcl}
\la f_n|\rho |f_n\ra&\!=\!&
\left[\left(\begin{array}{c}{a^n_m}^*\\1\end{array}\right)\otimes
\left(\begin{array}{c}{a^n_{m-1}}^*\\1\end{array}\right)\otimes
\cdots
\otimes\left(\begin{array}{c}{a^n_1}^*\\1\end{array}\right)\right]
(a^n_m\,\,\, 1)\otimes(a^n_{m-1}\,\, 1)\otimes
\cdots\otimes(a^n_1\,\,\, 1)\\[6mm]
&=&|e_{A_1},\ e_{A_2},\cdots,\, \ e_{A_m}\ra\la e_{A_1},\
e_{A_2},\cdots , \ e_{A_m}|.
\end{array}
$$

We can thus write $\rho$ as
$$\rho=\sum_{n=1}^N|\psi_n\ra\la \psi_n|\otimes
|\phi_n\ra\la \phi_n|\otimes \cdots \otimes |\omega_n\ra\la
\omega_n|\otimes |f_n\ra\la f_n|,
$$
where
$$
|\psi_n\ra=\left(\begin{array}{c}{a^n_m}^*\\
1\end{array}\right),~~~
|\phi_n\ra=\left(\begin{array}{c}{a^n_{m-1}}^*\\1\end{array}\right),~~~
\cdots,~~~
|\omega_n\ra=\left(\begin{array}{c}{a^n_1}^*\\1\end{array}\right).
$$
Because the local transformations are reversible, we can now apply
the inverse transformations and obtain a decomposition of the
initial state $\rho$ in a sum of projectors onto product vectors.
This proves the separability of $\rho$. \hfill $\Box$

By using Lemma 1, Theorem 1 and the method in \cite{KMN}, we can
generalize the results  to multipartite quantum systems in
${\Cb}^{K_1}_{\sc A_1} \otimes {\Cb}^{K_2}_{\sc A_2} \otimes
\cdots \otimes {\Cb}^{K_m}_{\sc A_m} \otimes {\Cb}^N_{\sc
A_{m+1}}$ with rank $N$. Let $|0_{A_i}\ra$, $|1_{A_i}\ra$,
$\cdots$, $|(K_i-1)_{A_i}\ra$, $i=1,2,\cdots, m$; and
$|0_{A_{m+1}}\ra\ , \cdots \ ,|(N-1)_{A_{m+1}}\ra$ be some local
bases of the sub-systems $A_i$, $i=1,2,\cdots, m , ~A_{m+1}$
respectively.

{\bf Lemma 2. }\ \  Every PPT state $\rho$ in ${\Cb}^{K_1}_{\sc
A_1} \otimes {\Cb}^{K_2}_{\sc A_2} \otimes \cdots \otimes
{\Cb}^{K_m}_{{\sc A}_m} \otimes {\Cb}^N_{{\sc A}_{m+1}}$ such that
$
r(\la {(K_1-1)}_{A_1}, \cdots,
{(K_m-1)}_{A_m}|\rho |{(K_1-1)}_{A_1}, \cdots,
{(K_m-1)}_{A_m}\ra)=r(\rho)=N,
$
can be transformed into the following canonical form by using a
reversible local operation: \be
\label{lemma2}\rho=\sqrt{F}T^{\dag}T \sqrt{F}, \ee where
$T=(D^1_{K_1-1}\,\,\cdots \,\,\ D^1_1\,\, \ I) \otimes
(D^2_{K_2-1}\,\,\cdots \,\,\ D^2_1\,\,\ I) \otimes \cdots \otimes
(D^m_{K_{m-1}}\,\,\cdots \,\,\ D^m_1\,\,\ I)$, $D^t_i$, $F$ and
the identity $I$ are $N\times N$ matrices acting on ${\Cb}_{{\sc
A}_{m+1}}^N$ and satisfy the following relations: $[D^t_{i_t},\
D^s_{j_s}]=[D^p_{i_p},\ {D^q_{j_q}}^{\dag}]=0$, and $F=F^{\dag}$ ,
$i_t,j_t=1,2,\cdots \ {K_t}$, $t,s,p,q=1,2,\cdots \ m$.

{\bf Proof.} We prove the lemma by induction on $m$. It is
already proved for the cases $m=1$ \cite{hlpre} and $m=2$
\cite{KMN}.

Now we consider the case of general $m$. Suppose that for the case
$m-1$ the result is correct. In the considered basis a density
matrix $\rho$ can be always written as: $S\times S$-partitioned
matrix, where $S={K_1}{K_2}\cdots {K_m}$ , with the $i$-row
$j$-column entry $E_{ij}$, $r(E_{S\times S})=N$.

The projection $ \la {(K_1-1)}_{A_1}|\rho|{(K_1-1)}_{A_1}\ra $
gives rise to a state $\tilde{\rho}=\la
{(K_1-1)}_{A_1}|\rho|{(K_1-1)}_{A_1}\ra$
 which is  a state in ${\Cb}^{K_2}_{\sc A_2} \otimes
\cdots \otimes{\Cb}^{K_m}_{{\sc A}_m} \otimes {\Cb}^N_{{\sc
A}_{m+1}}$ with $r(\tilde{\rho})=r(\rho)=N$.
  The fact that $\rho$ is PPT implies that $\tilde \rho$ is
also PPT, $\tilde{\rho}\ge 0$. By induction hypothesis , we have
\be\label{w1} \tilde{\rho}=\sqrt{F}{T_1}^{\dag}{T_1} \sqrt{F}, \ee
where $T_1=(D^{2}_{K_2-1}\,\,\cdots \,\,\ D^{2}_1\,\,\ I) \otimes
\cdots \otimes (D^{m}_{K_{m-1}}\,\,\cdots \,\,\ D^{m}_1\,\,\ I)$,
with $[D^{t}_{i_t},\ D^{s}_{j_s}]=[D^{p}_{i_p},\
{D^{q}_{j_q}}^{\dag}]=0$, $i_t,j_t=1,2,\cdots ,\ {K_t}$,
$t,s,p,q=2,\cdots, \ m$.

Similarly, if we consider the projection $\la
(K_2-1)_{A_2}|\rho|(K_2-1)_{A_2}\ra$, $\cdots$, $\la
(K_m-1)_{A_m}|\rho|(K_m-1)_{A_m}\ra$, by induction hypothesis, we
have $m-1$ relations like $(\ref{w1})$.  In fact, the projection
$\la (K_j-1)_{A_j}|\rho|(K_j-1)_{A_j}\ra$, $ j=2,\cdots, m$, is
\be\label{wj} {\rho}_j=\sqrt{F}{T_j}^{\dag}{T_j} \sqrt{F}, \ee
where
$$
\ba{l} T_j=(D^{1}_{K_1-1}\,\,\cdots \,\,\ D^{1}_1\,\,\ I) \otimes
\cdots
\otimes (D^{j-1}_{K_{j-1}-1}\,\,\cdots \,\,\ D^{j-1}_1\,\,\ I)\\[3mm]
~~~~~~~~~~~~~\otimes (D^{j+1}_{K_{j+1}-1}\,\,\cdots \,\,\
D^{j+1}_1\,\,\ I)\otimes \cdots \otimes (D^{m}_{K_{m-1}}\,\,\cdots
\,\,\ D^{m}_1\,\,\ I), \ea
$$
with $[D^{t}_{i_t},\ D^{s}_{j_s}]=[D^{p}_{i_p},\
{D^{q}_{j_q}}^{\dag}]=0$, $i_t,j_t=1,2,\cdots,{K_t}$,
$t,s,p,q=1,\cdots ,j-1, j+1 ,\cdots \ m$. Taking into account all
these projections and using the kernel vectors of $\rho$ we can
determine part of the entries $E_{ij}$. Then by using the PPT
property of the partial transpose of $\rho$ related to the
subsystems: the kernel vectors of $\rho$ are also the kernel
vectors of the partial transpose of $\rho$; if $E_{jj}$ is known,
the rest entries of $\rho$, $E_{ij}$, $i< j$, are also determined.
For the diagonal elements like $E_{11}$, we define
$\Delta=E_{11}-S^{\dag}S$, where $S=D^1_{K_1-1}D^2_{K_2-1}\cdots
D^m_{K_m-1}$, to determine $E_{11}$ by proving that $\Delta=0$. It
is straightforward to prove that there exist  matrices
$D^i_{s_i}$, $s_i=1,2,\cdots, K_i-1$, $i=1,2,\cdots, m$ satisfying
the relation $(\ref{lemma2})$. \hfill $\Box$

From Lemma 2 we have the following conclusion:

{\bf Theorem 2.}\ \ A PPT-state $\rho$ in ${\Cb}^{K_1}_{\sc A_1}
\otimes {\Cb}^{K_2}_{\sc A_2} \otimes \cdots \otimes
{\Cb}^{K_m}_{{\sc A}_m} \otimes {\Cb}^N_{{\sc A}_{m+1}}$  with
$r(\rho)=N$ is separable if there exists a product basis
$|e_{A_1},\ e_{A_2},\cdots , \ e_{A_m}\ra$ such that
$$r(\la
e_{A_1},\ e_{A_2},\cdots, \ e_{A_m}|\rho|e_{A_1},\ e_{A_2},
\cdots, \ e_{A_m} \ra)=N$$.

 We have derived a canonical form of
PPT states in ${\Cb}^{K_1} \otimes {\Cb}^{K_2} \otimes \cdots
\otimes {\Cb}^{K_m} \otimes {\Cb}^N$ composite quantum systems
with rank $N$, together with a sufficient separability criterion
from this canonical form. Generally the separability criterion we
can deduce here is weaker than in the cases of ${\Cb}^{2} \otimes
{\Cb}^{2}\otimes {\Cb}^N$ and ${\Cb}^{2} \otimes {\Cb}^{3}\otimes
{\Cb}^N$, as the PPT criterion is only sufficient and necessary
for the separability of bipartite states on ${\Cb}^{2} \otimes
{\Cb}^{2}$ and ${\Cb}^{2} \otimes {\Cb}^{3}$. Besides the
discussions of separability criterion, the canonical
representation of PPT states can shade light on studying the
structure of bound entangled states.
One can check if these PPT states
are bound entangled by checking whether they are
entangled or not.


\vskip 8mm
\begin{thebibliography}{20}

\bibitem{bennet98} C. H. Bennet, {\em Phys. Scr. T} {\bf 76}, 210 (1998).

\bibitem{book} M. Nielsen and I. Chuang, Quantum Computation and
Quantum Information (Cambridge University Press, Cambridge,
England, 2000).

\bibitem{peresbook} A. Peres, ``Quantum Theory: Concepts and Methods'',
Kluwer Academic Publishers (1995).

\bibitem{22n}
S. Karnas and  M. Lewenstein, Phys. Rev. A 64, 042313 (2001).

\bibitem{23n}
S.M. Fei, X.H. Gao, X.H. Wang, Z.X. Wang and K. Wu, { Phys. Rev.
A} {68} (2003)022315.

\bibitem{KMN}
S.M. Fei, X.H. Wang, Z.X. Wang and K. Wu, {\it On PPT States in
${\Cb}^K \otimes {\Cb}^M\otimes {\Cb}^N$ Composite Quantum
Systems}, to be appear in Commun.Theor. Phys. (2004).

\bibitem{222n}
S.M. Fei, X.H. Gao, X.H. Wang, Z.X. Wang and K. Wu, Commun. Theor.
Phys. 40 (2003) 515-518.

\bibitem{hlpre}
S.M. Fei, X.H. Gao, X.H. Wang, Z.X. Wang and K. Wu, {\sf Int. J.
Quant. Inform.}, {\bf 1}(2003)37-49.


\end{thebibliography}
\end{document}